\documentclass[12pt]{iopart}
\usepackage{graphics,graphicx,marvosym}
\usepackage{dcolumn,bm}
\usepackage{mathbbol}

\begin{document}
\title[]{Thermal Rectification and Negative Differential Thermal Resistance
in a driven two segment classical Heisenberg chain}
\author{Debarshee  Bagchi}
\address{Theoretical Condensed Matter Physics Division, 
Saha Institute of Nuclear Physics, 1/AF Bidhannagar, Kolkata 700064, India}
\eads{debarshee.bagchi@saha.ac.in}
\date{\today}

\begin{abstract}
We investigate thermal transport in a two segment classical Heisenberg spin chain with nearest
neighbor interaction and in presence of external magnetic field using computer simulation. The
system is thermally driven by heat baths attached at the two ends and transport properties are
studied using an energy
conserving dynamics. We demonstrate that by properly tuning the parameters
thermal rectification can be achieved - the system behaves as a good conductor of heat along one
direction but becomes a bad conductor when the thermal gradient is reversed and crucially depends
on nonlinearity and spatial asymmetry. Moreover, suitable tuning of the system parameters gives
rise to the counterintuitive and technologically important feature known as the negative differential
thermal resistance (NDTR). We find that the crucial factor responsible for the emergence of NDTR is
a suitable mechanism to impede the current in the bulk of the system.
\end{abstract}

\maketitle

\section{Introduction}

%General Introduction
Thermal management in low dimensional mesoscopic systems is an active topic of research at present.
The quest to manipulate and control heat current, just as one can do with electrical current in
electronic devices, has given birth to an altogether new branch of study namely the {\it phononics}
\cite{phononics,control}. Needless to say, these studies are important not only in understanding
the principles of low dimensional thermal transport, for example, necessary and sufficient
condition for Fourier law \cite{Lebo,Lepri,ADrev}, but also have immense technological application
in today's world.
Phononics deals with the manipulation of thermally excited phonons in the system and is now quite a
mature field of research. Designs for many useful thermal devices have been proposed in recent times
such as the thermal rectifier \cite{diode1,diode2}, transistor \cite{transistor}, logic gates \cite{gate},
memory elements \cite{memory}, current limiter and constant current source \cite{limiter}.
In fact, very recently a thermal rectifier \cite{diode} and a thermal wave guide \cite{wguide},
both using carbon and boron nitride nanotubes, have been successfully fabricated in the laboratory.
Apart from phonons, spin waves (magnons) in magnetic systems are also known to be an efficient
mode of energy transport for quite some time now \cite{magnon_transport}. Although considerable
progress has been achieved in understanding magnon assisted thermal transport \cite{magnon_review},
a lot of effort still needs to be devoted before this can be utilized in real thermal devices.

%Our Work
%Review of old works
The Heisenberg model \cite{fisher,joyce} is a paradigmatic model for magnetic insulators.
Thermal transport properties of the one dimensional classical Heisenberg model have been
studied in recent times and it is now known that in the thermodynamic limit heat transport
in this model obeys Fourier law (diffusive transport of energy) for all temperature and
magnetic field \cite{savin,our}. This diffusive energy transport is attributed to the
nonlinear spin wave interactions in this model which cause spin waves to scatter \cite{savin}.
However, for finite systems there can be a crossover from ballistic to diffusive behavior
which crucially depends on the temperature  \cite{our} and also on other system parameters
such as an external magnetic field. This is due to the fact that at very low temperature or
very high magnetic field the entire system becomes correlated and heat energy can pass from
the hotter to the colder end without being scattered.

In this paper we study the thermal transport in a two segment classical Heisenberg spin chain.
The two segments are connected to each other by a link and external magnetic fields act on
the spins in the chain. Heat baths are attached to the two ends of the system and an energy
current flows through the system in response to the imposed thermal gradient. It is
found that the thermal current can be rectified by suitably tuning the system parameters. Thus
heat current can preferentially flow through the system along one direction while it is inhibited
in the opposite direction i.e., when the thermal gradient is reversed.

Over the last decade, thermal rectification has been intensively investigated in a large number
of theoretical as well as experimental works and in a variety of systems; for a recent review see
\cite{rect_review}. Thermal rectification has been observed in many classical nonlinear asymmetric
lattices in one dimension with different forms of interaction potentials e.g., Morse \cite{diode1,morse},
FK \cite{diode2,FK-FK}, FK-FPU \cite{FK-FPU}, graded mass harmonic systems \cite{graded1,graded2}
to name a few. Studies of thermal rectification in two \cite{2D} and three \cite{3D} dimensions have
also been done recently.

Another counterintuitive feature that emerges in some of these systems is the negative differential
thermal resistance (NDTR) \cite{diode2}. In the NDTR regime thermal current through a driven system
is found to decrease as the imposed thermal gradient is increased. Although a lot of work
\cite{origin,ballistic_yes,ballistic_NO,transition,scaling} has been done to figure out the
criteria responsible for the origin of NDTR, a comprehensive understanding is still lacking.
This is however highly desirable now, since NDTR is a promising feature and is believed to be
crucial in the functioning of thermal devices, such as thermal transistors \cite{transistor}
and thermal logic gates \cite{gate}.

In this paper we focus particularly on these two features - thermal rectification and negative
differential thermal resistance. Rectification and NDTR have been observed using computer
simulation for the classical two dimensional Ising spin system \cite{ising}. Although
rectification was attributed to the difference in the temperature dependence of thermal conductivity
of the two segments, the origin of the intriguing NDTR effect was not discussed. Also there are certain
differences between the Ising system and our chosen spin model. Firstly, Ising model is a discrete spin
model and, although simple, it is not very realistic. A more realistic spin model is the Heisenberg model
with continuous spin degree of freedom.
Secondly, the $2$d Ising model has a phase transition at a finite temperature whereas, the $1$d Heisenberg
model does not have a phase transition at any finite temperature. However thermal transport is found to be
diffusive (obeys Fourier law) for both the models. Since an exact analytical treatment is generally not
possible for most of the models mentioned above, an effective alternative is to investigate different
generic models using numerical simulation. In the present work, we undertake such a numerical study of
the above-mentioned features in a driven classical Heisenberg spin system.

%Organisation
The organization of the paper is as follows. We define the two segment classical Heisenberg
spin chain in detail in Sec. \ref{model}. The simulation of the model has been performed
using the discrete time odd even (DTOE) dynamics. The numerical implementation of this
dynamics is briefly discussed in Sec. \ref{scheme}. Thereafter in Sec. \ref{results}, we
present our numerical results and demonstrate the existence of thermal rectification and
negative differential thermal resistance in this system.
We study dependencies of system parameters on these features and also investigate the underlying
physical mechanism for the emergence of the NDTR regime. Finally, we conclude by summarizing
our main results and with a discussion in Sec. \ref{summary}.

\section{Model}
\label{model}
\begin{figure}[htbp]
\begin{center}
\includegraphics[width=8.5cm,height=2.65cm,angle=0]{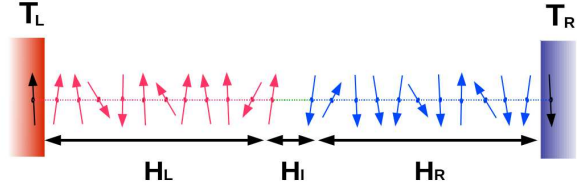}
\end{center}
\caption{(Color online) Schematic diagram of the two segment classical Heisenberg model.}
\label{fig:model}
\end{figure}

A schematic diagram of the model studied here is shown in Fig. \ref{fig:model}. Consider two
one-dimensional segments of spins connected to each other by a link.  The spins $\vec S^L_i$
($\vec S^R_i$) are conventional  classical Heisenberg spins  belonging to the left (right)
segment of the chain and $i$ is the site index which runs from $1 \le i \le N_L (N_R)$. Each
spin on the left (right) segment interacts with an external magnetic field $\vec h_i^L$
($\vec h_i^R$). The Hamiltonian of the system is given as,

\begin{equation}
 \mathcal{H} = \mathcal{H}_L + \mathcal{H}_I + \mathcal{H}_R
\label{ham}
\end{equation}
and the interaction of the spins in the left and the right segments are

\begin{eqnarray}
\mathcal{H}_L &=& -K_L \sum_{i=1}^{N_L-1}\vec{ S^L}_i\cdot \vec{S^L}_{i+1} - \sum_{i=1}^{N_L}\vec h^L_i \cdot \vec S^L_i \cr
\mathcal{H}_R &=& -K_R \sum_{i=1}^{N_R-1}\vec{ S^R}_i\cdot \vec{S^R}_{i+1} - \sum_{i=1}^{N_R}\vec h^R_i \cdot \vec S^R_i,
\label{H_LR}
\end{eqnarray}
where the $K$'s are the interaction strengths which is ferromagnetic for coupling 
$K > 0$ and anti-ferromagnetic for $K < 0$. The interaction of the last spin of the
left segment $i = N_L$ and the first spin of the right segment $i = 1$ is chosen
to be of the ferromagnetic Ising-Heisenberg form \cite{iheis}

\begin{equation}
 \mathcal{H}_{I} = -K_I \left[\lambda \,(S^{x}_i S^{x}_j + S^{y}_i S^{y}_j) + S^{z}_i S^{z}_j\right],
\label{H_I}
\end{equation}
where $\vec S_i = \vec S^L_{N_L}$ and $\vec S_j = \vec S^R_1$ which defines the link connecting the two
segments. For $\lambda = 1$, we refer to the interaction at the interface as the (isotropic) ``Heisenberg''
type, whereas, for $\lambda = 0$, the ``Ising'' type. For intermediate values of $\lambda$ we have
(anisotropic) ``XXZ'' type interaction at the interface. Note that the spins are free to fluctuate
in all the three directions for all $\lambda$.

The evolution equation for the spin vectors in the two segments can be written as,
\begin{equation}
\frac d{dt} {\vec{ S}_i} = \vec{S}_i \times \vec{B}_i,
\label{eom}
\end{equation}
where $\vec{B}_i = (K_i^-\vec{S}_{i-1} + K_i^+\vec{S}_{i+1}) + \vec h_i$ is the local molecular
field experienced by the spin  $\vec S_i$; $K_i^-$ and $K_i^+$ are the interactions of $\vec S_i$
with $\vec S_{i-1}$ and $\vec S_{i+1}$ respectively and will be $K_L$, $K_R$ or $K_I$  depending on
the segment in which the spins $\vec S_{i-1}$, $\vec S_i$ and $\vec S_{i+1}$ belong to; $\vec h_i$
is the magnetic field acting on the $i-$th spin. For the two spins at the interface the equation
of motion can be written down analogously.

This two segment system is thermally driven by two heat baths attached to the two ends.
This is numerically implemented by introducing two additional spins at sites $i = 0$ on
the left segment and $i = N_R + 1$ on the right segment. The bonds between the pairs of
spins ($\vec S^L_0, \vec S^L_1$) and  ($\vec S^R_{N_R}, \vec S^R_{N_R +1}$) at two ends
of the system behave as stochastic thermal baths  \cite{our}. The interaction strength
of the bath spins with the system is taken to be $K_b$, and therefore energy of both the
baths is bounded in the range $(-K_b, K_b)$ and has a Boltzmann distribution. Thus the
left and right baths are in equilibrium at their respective temperatures, $T_L$ and $T_R$
with average energies
$E_L  = - K_b ~ \mathcal{L}(K_b/T_L)$ and $E_R = - K_b~ \mathcal{L}(K_b/T_R)$,
$\mathcal{L}(x)$ being the standard Lang\'{e}vin function.
Thus one can set the two baths at a fixed average energies (or fixed temperatures) and an
energy current flows through the system if $T_L \neq T_R$. In the steady state a uniform
current (independent of the site index $i$) transports thermal energy from the hotter to
the colder end of this composite system.

\section{Numerical Scheme}
\label{scheme}

We investigate transport properties of this composite two segment Heisenberg model
by numerically computing the steady state quantities, such as, currents, energy
profiles etc. using the energy conserving DTOE dynamics \cite{our,sdiff}. The DTOE dynamics
updates spins belonging to the odd and even sub-lattices alternately using a spin
precessional dynamics
\begin{equation}
\vec{S}_{i,t+1} = \left[\vec{S} \cos \phi + (\vec{S} \times \hat{B}) \sin \phi + (\vec{S}\cdotp\hat{B})\hat{B}(1 - \cos \phi) \right]_{i,t}
\label{precess}
\end{equation}
where $\hat{B}_i = \vec{B}_i/|\vec{B}_i|$, $\phi_i = |\vec{B}_i| \Delta t$ and $\Delta t$
is the integration time step. The above formula is sometimes referred to as the
\textit{rotation formula} and holds for any finite rotation \cite{goldstein}.
Note that Eq. \ref{precess} reduces to the equation of motion Eq. \ref{eom} in the limit
$\Delta t \to 0$.
The bath spins are refreshed by drawing the bond energies between the spins ($\vec S^L_0, \vec S^L_1$) and
($\vec S^R_{N_R}, \vec S^R_{N_R +1}$) from their respective Boltzmann distribution consistent with the
temperature of the left and right bath temperatures. The left bath (at $i = 0$) is updated along with the
even spins and the right bath is updated along with the even or odd spins depending on whether $i = N_R + 1$
is even or odd.

%\subsection{Thermal Current}
The DTOE dynamics alternately updates only half of the spins (odd or even) but all the bond energies
are updated simultaneously. So the energy of the $i$-th bond $\varepsilon^o_i$ measured immediately
after the update of odd spins is not equal to its bond energy $\varepsilon^e_i$ measured after the
subsequent update of even spins, where we define the energy density
$\varepsilon_i = - \vec S_i\cdot \left[K_i^ + \vec S_{i+1} + \vec h_i\right] $.
The difference $\varepsilon^e_i - \varepsilon^o_i$ is the measure of the heat energy passing through
the $i$-th bond in time $\Delta t$. Therefore the current $J$ (rate of flow of energy) in the steady
state is given by

\begin{equation}
J = \langle \varepsilon^e_i - \varepsilon^o_i  \rangle/\Delta t.
\label{J}
\end{equation}
Note that Eq. (\ref{J}) can be shown to be consistent with the definition of current obtained
using the continuity equation \cite{our}. Numerically however, the current $J$ can be computed as
\begin{equation}
J = - K_i^+ \langle \left(\vec S_i \cdot \vec S_{i+1} \right)^e - \left(\vec S_i \cdot \vec S_{i+1} \right)^o  \rangle/\Delta t,
\end{equation}
since the magnetic field term cancels out in the above expression.
The energy profile $E_i$ for sites in the two segments is computed as
\begin{equation}
E_i =  - \left \langle \frac12 \vec S_i \cdot (K_i^- \vec S_{i-1} + K_i^+ \vec S_{i+1}) + \vec h_i \cdot \vec S_i \right \rangle,
\label{eprof}
\end{equation}
For the left and the right boundaries (bath sites) energy is calculated as,
\begin{equation}
E_L = -K_b \langle \vec S_0^L \cdot \vec S_1^L \rangle \hskip1cm
E_R = -K_b \langle \vec S_{N_R}^R \cdot \vec S_{N_R + 1}^R \rangle.
\end{equation}
In the following we present our numerical results obtained using the DTOE dynamics for the
thermally driven two segment classical Heisenberg model.

\section{Numerical Results}
\label{results}

We study the two segment system with boundary temperatures $T_L = T_0 (1+\Delta)$ and
$T_R = T_0 (1-\Delta)$; the average temperature of the system can be taken as
$\frac 12(T_L + T_R) = T_0$. We set the segment sizes $N_L = N_R = N$, and the parameters
are set as $K_b = K_L = 1$, $K_R = K$, $\vec h_i^L = (0,0,1)$ $\vec h_i^R = (0,0,h)$. Thus
the parameters we can manipulate reduce to $\Delta, K, h$ and the interface parameters
$(K_I, \lambda)$, all of which are kept restricted in the range $(0,1)$ unless mentioned
otherwise.
The time step $\Delta t$ is chosen relatively larger since a larger $\Delta t$ ensures faster
equilibration of the system and because energy conservation is maintained for any finite
$\Delta t$ \cite{our}. Also it can be shown that the final stationary state is the same for
all choices of $\Delta t$ \cite{our}. In the following we set $\Delta t = 2.0$ for all our
simulations.
Starting from a random initial configuration we evolve the spins using the DTOE dynamics.
Once a nonequilibrium stationary state is reached, we compute various quantities, such as,
the thermal current and the energy profiles for different values of the above mentioned
parameters, temperature and system size.

\subsection{Thermal rectification}
\begin{figure}[htbp]
\begin{center}
\hskip-0.5cm
\includegraphics[width=3.35cm,angle=-90]{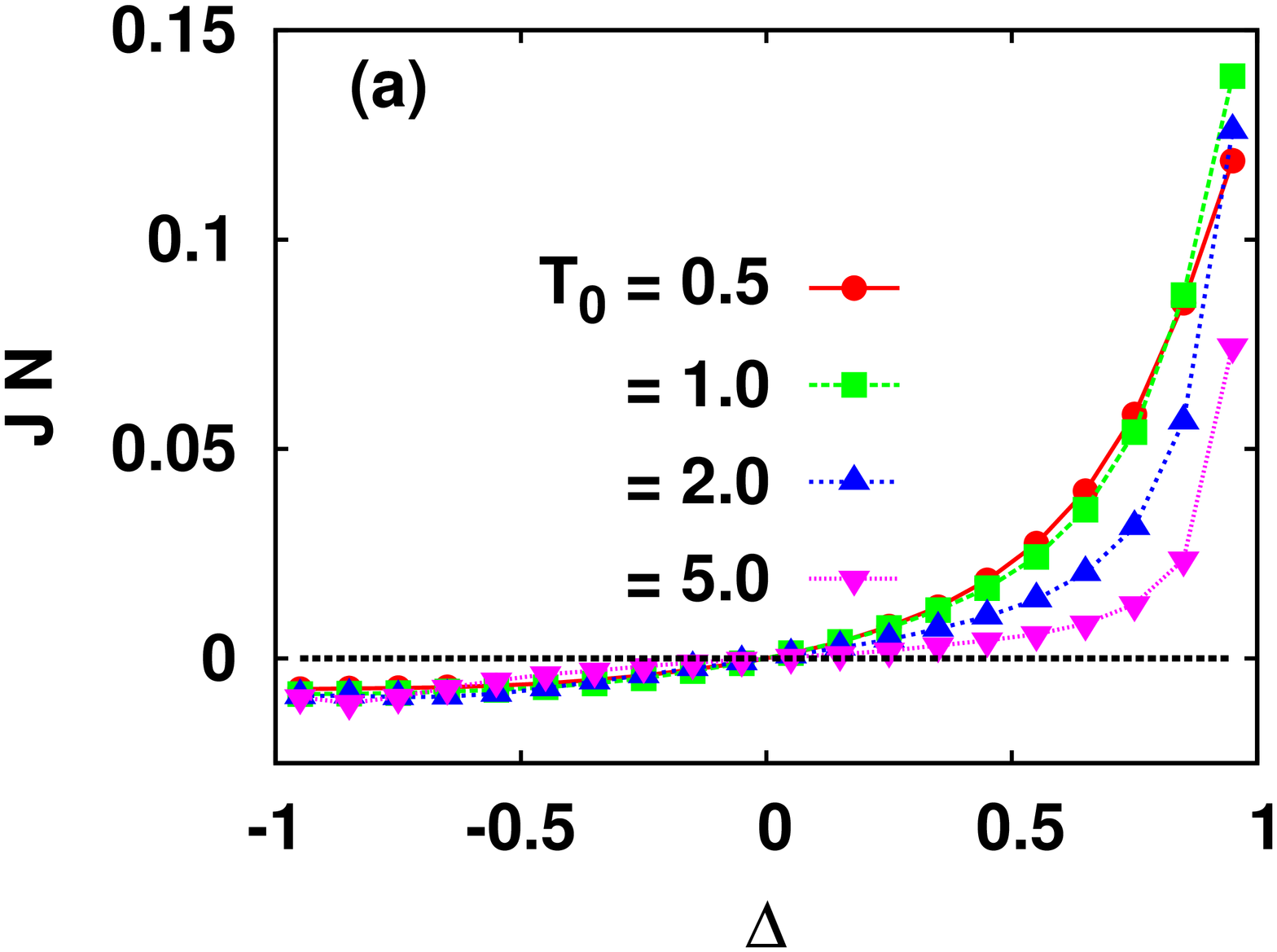}
\hskip-0.5cm
\includegraphics[width=3.35cm,angle=-90]{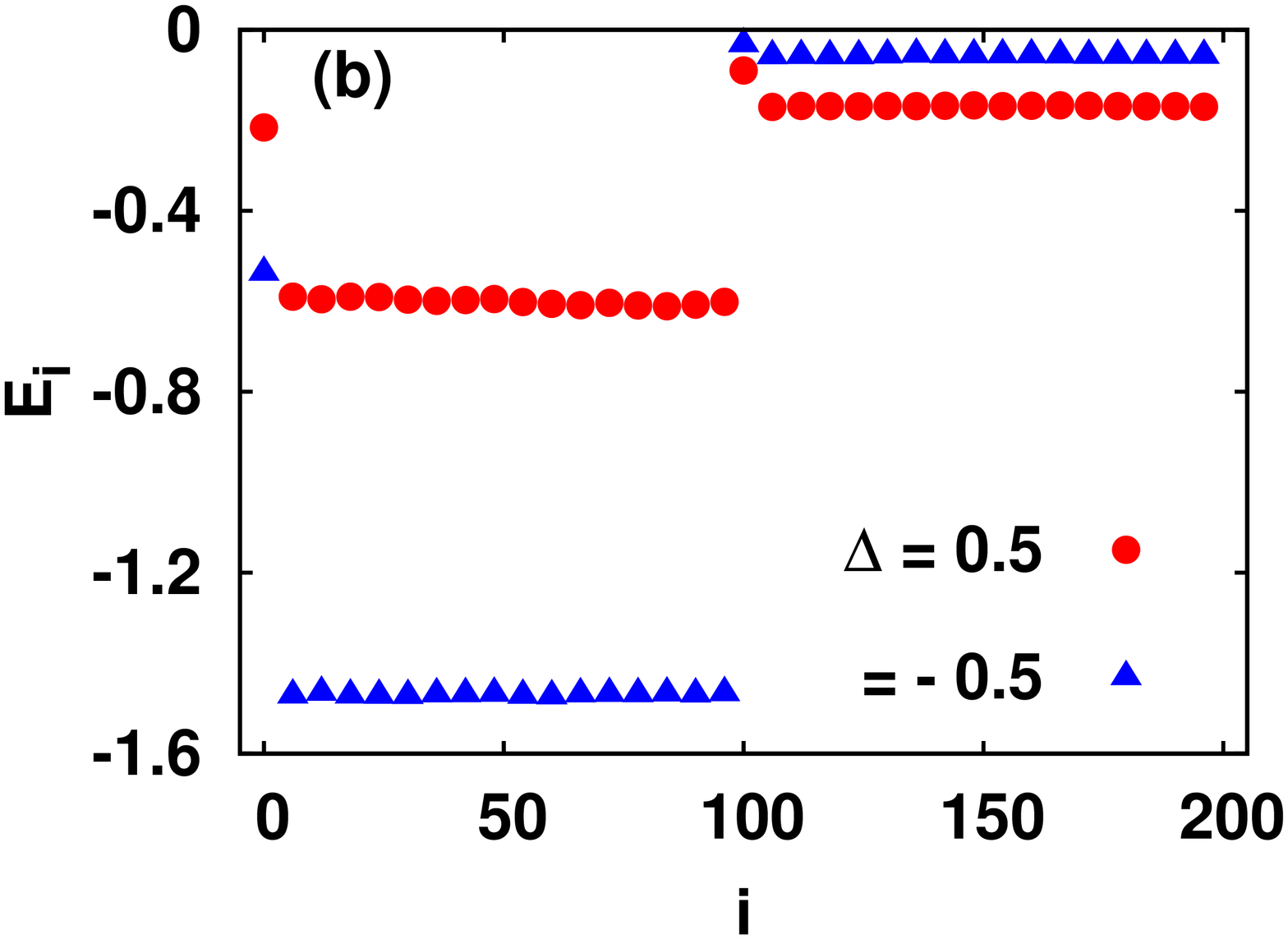} 
\end{center}
\caption{(Color online) (a) Variation of the thermal current $J$ with $\Delta$ for different
average temperatures $T_0 = 0.5, 1.0, 2.0$ and $5.0$. (b) Energy profile of the system for
$\Delta = 0.5$ and $-0.5$; average temperature $T_0=1.0$. The parameters used for both the
figures are $\lambda = 1$, $K=0.5$, $K_I=0.05$, $h = 0.1$ and segment size $N=100$.
}
\label{fig:Diode}
\end{figure}

First, we consider the system with $\lambda = 1$ (``Heisenberg'' type interaction at the
interface) and study the thermal current through the system for different values of the parameter
$-1 < \Delta < 1$. A positive $\Delta$ implies that $T_L > T_R$, whereas, the same $\Delta$
with a negative sign implies that the heat baths  have been swapped between the two ends.
We refer to $\Delta > 0$ as the forward bias and $\Delta < 0$ as the backward bias. Thus
the entire system is now a Heisenberg spin chain with spatial asymmetry due to dissimilar
spin-spin coupling strengths and magnetic fields in the two segments. The variation of the
thermal current with the bias $\Delta$ for different values of average temperature $T_0$ is
shown in Fig. \ref{fig:Diode}a. We find that the thermal current is considerably larger for
$\Delta > 0$ whereas, for the same value of $\Delta$ but with the baths interchanged, the
current through the system is smaller. Thus, this nonlinear asymmetric two segment system
behaves as a thermal rectifier i.e., it acts as a good conductor of heat in one direction
and as a bad conductor in the reverse direction. Evidently this rectification effect is more
pronounced at lower temperatures as can be seen in Fig. \ref{fig:Diode}a.

The steady state energy profiles of this two segment system is displayed in Fig. \ref{fig:Diode}b.
for $\Delta = 0.5$ and $-0.5$. It can be seen that the energy profile is almost flat implying that
the system is near the ballistic regime, and there is an energy jump at the interface. This energy
discontinuity is due to the interface thermal resistance, often referred to as the
{\it Kapitza resistance} \cite{kapitza}. The current that flows in the system depends essentially on
two factors namely the imposed bias and the interface resistance. The current increases as the bias
is increased but decreases if the interface resistance is high which diminishes the current carrying
capacity of the lattice i.e., its conductivity drops.
It can be seen from Fig. \ref{fig:Diode}b that the interface resistance is comparatively larger
for $\Delta = -0.5$ than for $\Delta = 0.5$. This disparity in the interface resistance along the
forward and the backward direction for the same bias magnitude $|\Delta|$ results in the rectification
of the thermal current. Thus in such nonlinear systems, the interface resistance is a function of the
imposed driving field and has unequal values in the forward and the backward direction due to the spatial
asymmetry of the lattice.

\begin{figure}[htbp]
\begin{center}
\hskip-0.25cm
\includegraphics[width=3.2cm,angle=-90]{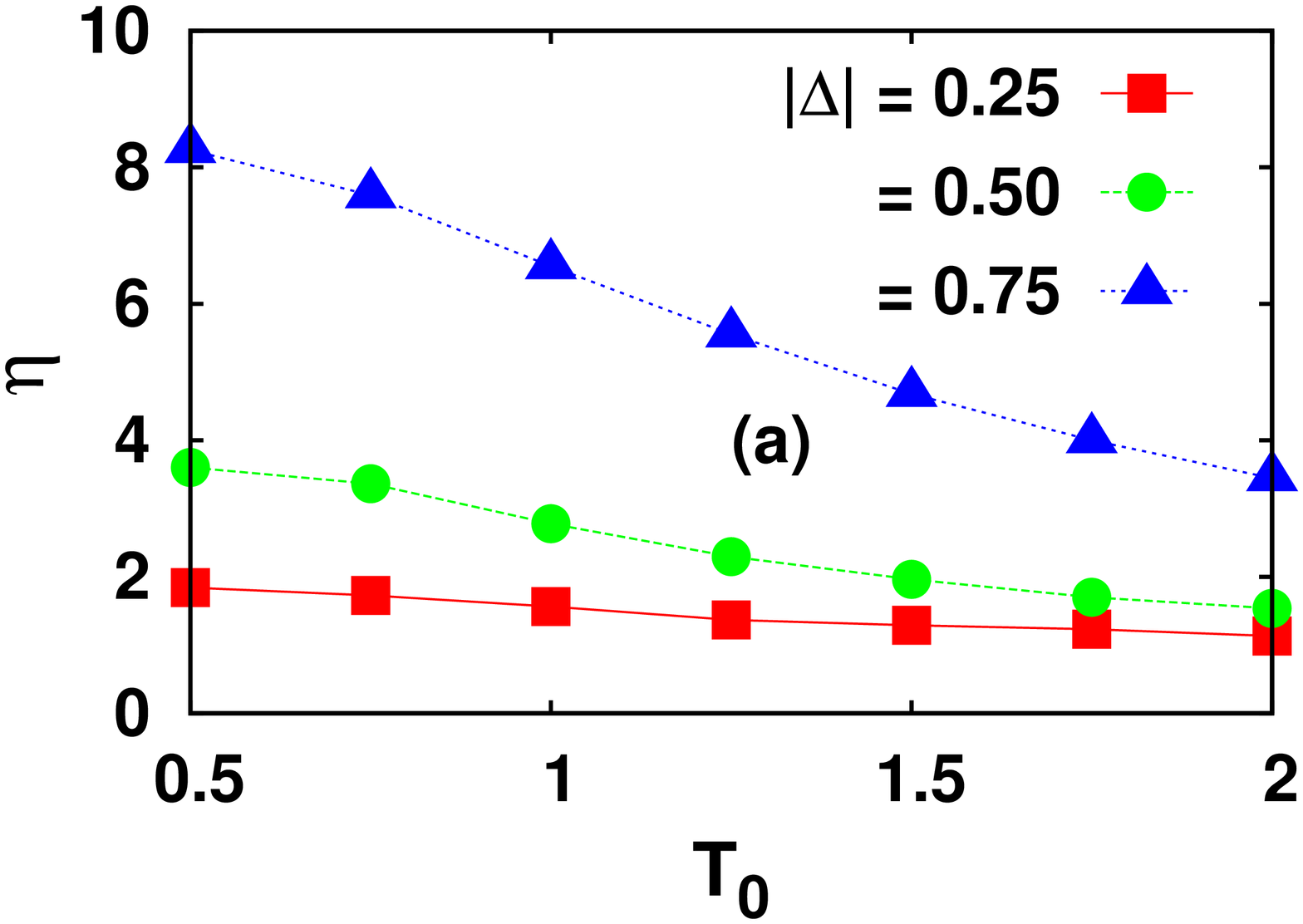}\hskip-0.25cm
\includegraphics[width=3.2cm,angle=-90]{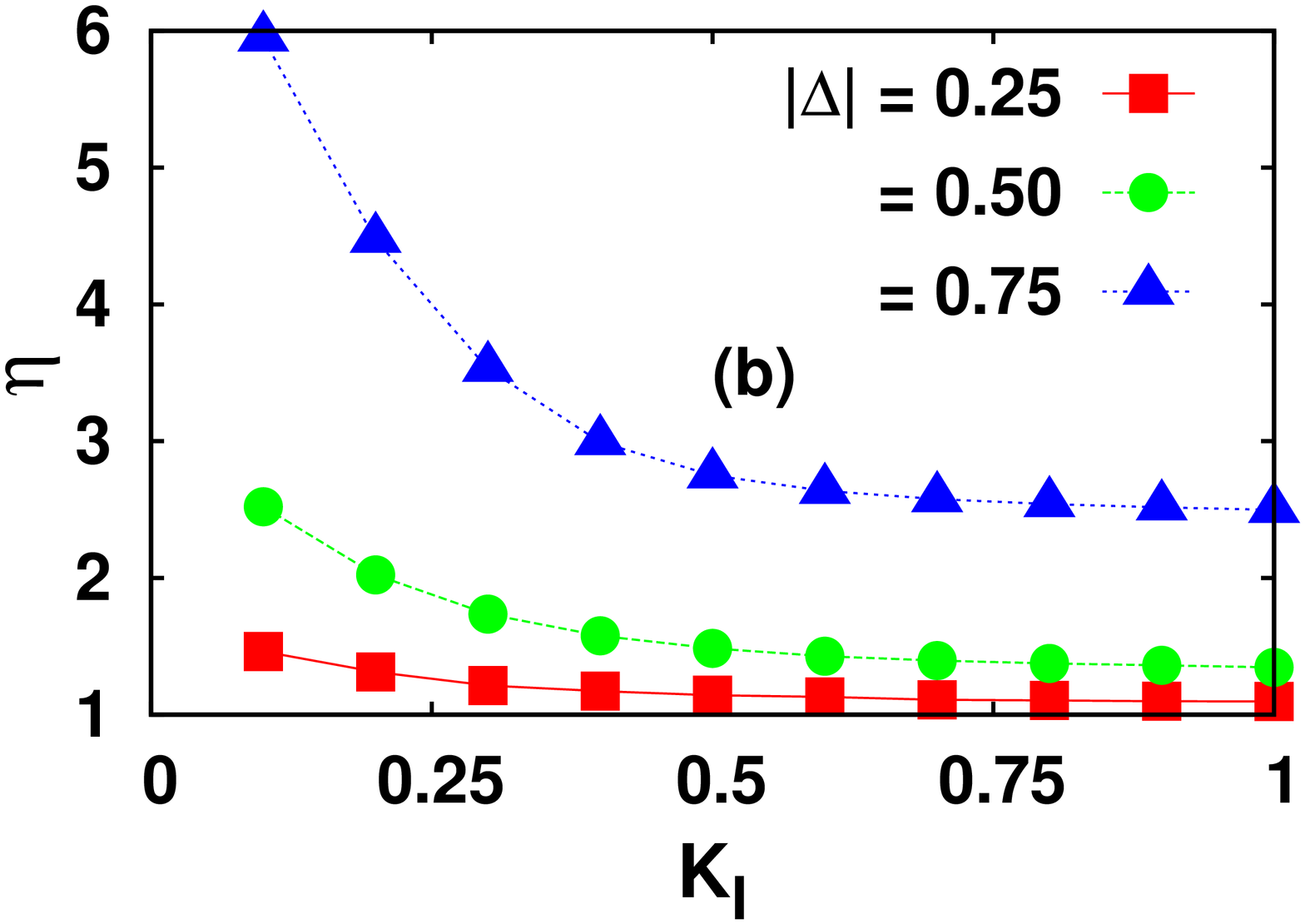}\\
\hskip-0.25cm
\includegraphics[width=3.2cm,angle=-90]{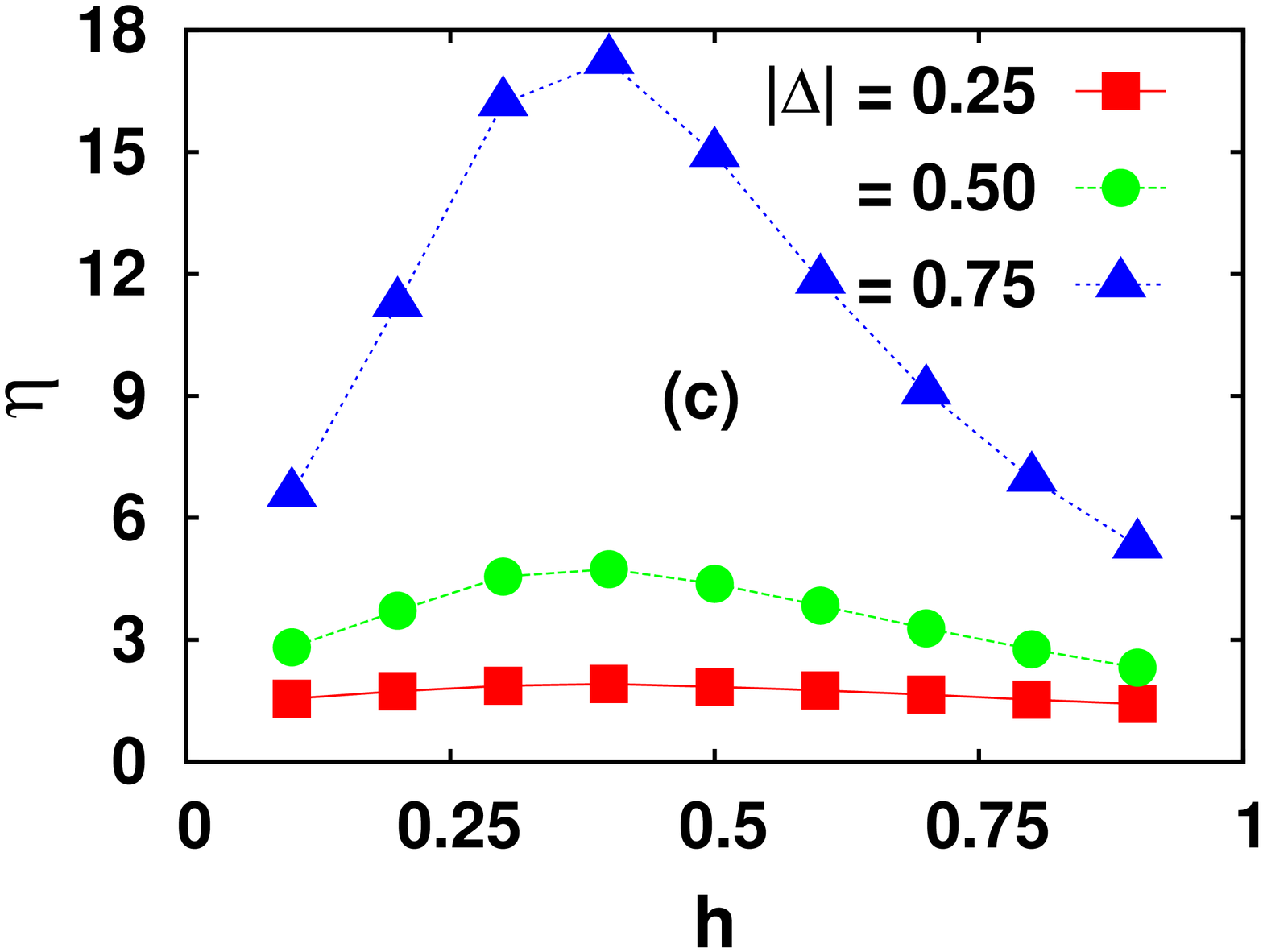}\hskip-0.25cm
\includegraphics[width=3.2cm,angle=-90]{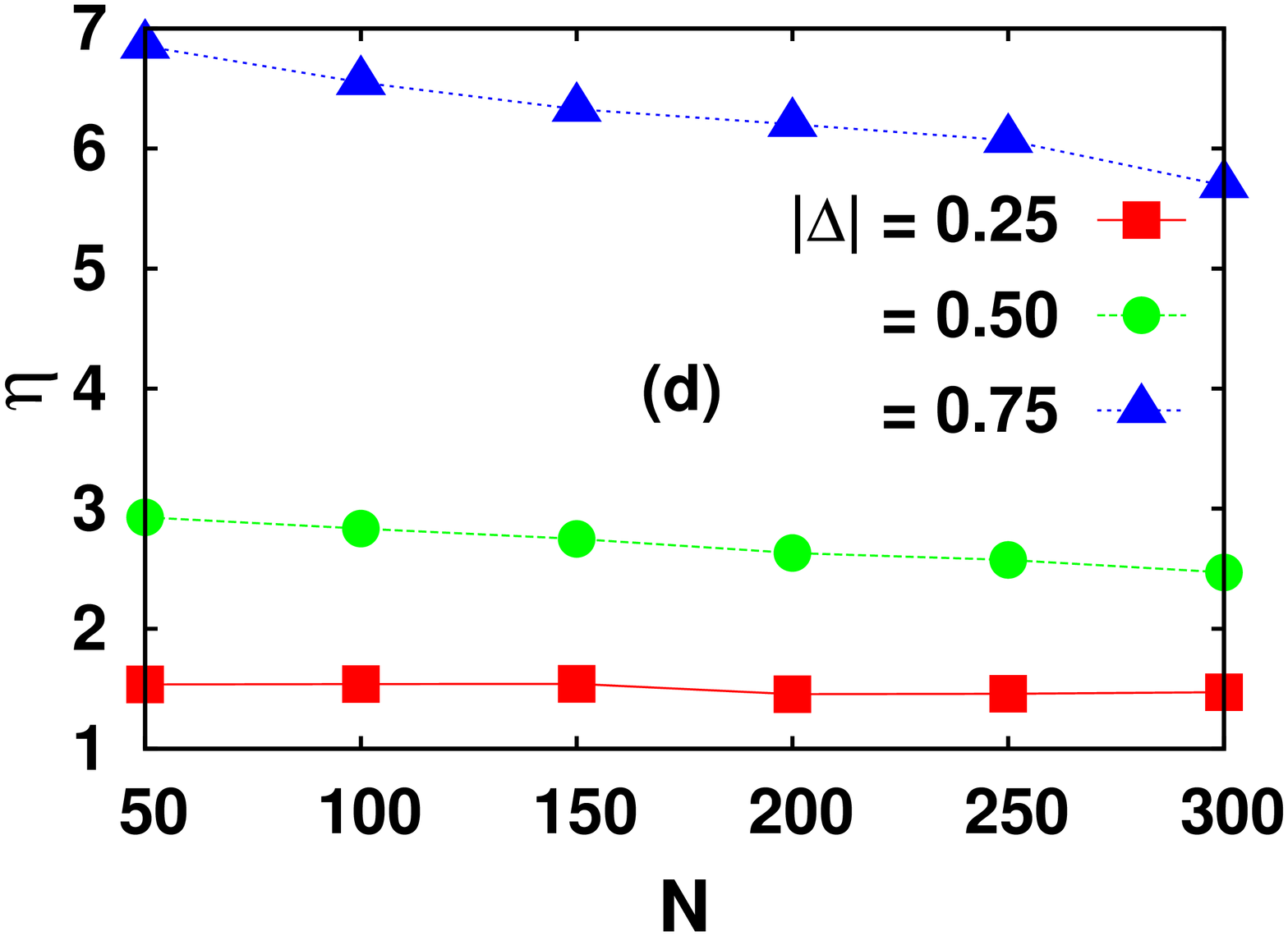}
\end{center}
\caption{(Color online) 
The variation of rectification efficiency $\eta$ with (a) average temperature $T_0$
(b) interface coupling $K_I$ (c) magnetic field strength $h$ and (d) segment size $N$ 
is shown for $|\Delta| = 0.25, 0.5$, and $0.75$. The values of the parameters are
chosen as $T_0 = 1.0$, $K_I = 0.05$, $h = 0.1$, $K = 0.5$ and $N = 100$. In each plot
only one of the above parameters is varied keeping all other values the same.
}
\label{fig:DiodePara}
\end{figure}

To quantify the amount of rectification, we compute the rectification efficiency defined
as  $\eta = |J_+/J_-|$ where both the forward and the backward currents, $J_+$ and $J_-$,
are computed for the same $|\Delta|$.  The variation of the rectification efficiency for
$|\Delta| = 0.25, 0.5$ and $0.75$ is shown in Fig. \ref{fig:DiodePara} for different values
of average temperature $T_0$, magnetic field $h$, interface coupling $K_I$ and segment size $N$.
The rectification is found to decrease as the temperature, interface coupling strength and the
system size increases. With magnetic field, the efficiency has a non-monotonic dependence in the
range $0 < h < 1$. It first increases and attains a maximum value corresponding to the field value
$h^*$ (say) and decreases thereafter decreases in the range $h^* < h < 1$. From the temperature and
the system size data it is clear that rectification is more when the system is closer to the ballistic
regime (lower temperature and smaller system size). As the magnetic field increases from zero, the
system moves closer to the ballistic regime and thus efficiency $\eta$ increases. However, as $h$
approaches unity, there is a drop in the efficiency since the asymmetry of the system is gradually lost.
Thus the non-monotonicity of the $\eta \sim h$ curve is related to this interplay between ballistic-diffusive
transport processes and asymmetry of the lattice. The efficiency attains a maximum value when
both these factors have an optimum value. If $h$ is increased beyond unity the rectification will again
start to increase due to spatial asymmetry. Since here rectification can be controlled externally by
tuning the magnetic fields, one can achieve quite large rectification efficiency in this system.

\begin{figure}[htbp]
\begin{center}
\hskip-0.55cm
\includegraphics[width=3.4cm,angle=-90]{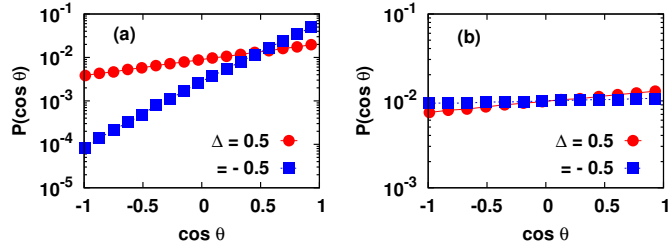}
\end{center}
\caption{(Color online) Semi-log plot of the distribution of $\cos \theta$ (or the $\hat z$
component of the spin) of spins (a) $\vec S^L_N$ and (b) $\vec S^R_1$, for forward and backward bias.
Here $\lambda = 1$, $|\Delta| = 0.5$ and $T_0 = 1.0$; other parameters are the same as in
Fig. \ref{fig:Diode}.
}
\label{fig:Pcos}
\end{figure}

Microscopically, if one looks at the two spins at the interface, namely  $\vec S^L_N$ and $\vec S^R_1$,
it is observed that the spins have unequal rotational \textit{stiffness} for the forward and backward
bias. By stiffness we mean the extent of rotation that is allowed for a particular spin about the
$\hat z$-axis i.e., the angle $\theta$ which is in the range $0 \le \theta \le \pi$. If a spin $\vec S_i$
can rotate completely freely then $S_i^z = \cos\theta$ should be a uniform distribution in the range
$-1\le \cos\theta \le 1$ with zero mean. However, if the spin is constrained to rotate within a restricted
angle then the mean $\langle \cos\theta \rangle$ is nonzero. For such a case, the larger the magnitude of
$\langle \cos \theta \rangle$ the more is the stiffness of the spin and less will be the current that passes
through the $i$-th site. In Fig. \ref{fig:Pcos}, we show the steady state distribution $P(\cos \theta)$ for
the two spins at the interface for $|\Delta| = 0.5$, both in the forward and backward bias.

We find that the distribution for the spin on the right segment $\vec S^R_1$ does not change much in
the forward and the backward bias. However the distribution for the left spin $\vec S^L_N$ changes
appreciably and the ratio $\langle \cos\theta \rangle_-/ \langle \cos\theta \rangle_+ \approx 2.5$.
Although this is rather crude, it gives a fairly good estimate of the amount of rectification achieved
from the system for the given set of parameter values as presented in  Fig. \ref{fig:Diode}a
($J_+/J_- \approx 2.8$ for $|\Delta| = 0.5$ at $T_0 = 1.0$). Note that, this unequal spin stiffness is
also the reason for the unequal energy discontinuity at the interface for the forward and backward
bias, as shown in Fig. \ref{fig:Diode}b. From the numerical data we estimate the ratio of the interface
energy jumps $\Delta E_-/\Delta E_+ \approx 3.0$. The left segment's stiffness dominates over that of the
right segment because of the higher values of interaction strength and magnetic field in the former. This 
disparity in spin stiffness at the interface inhibits the current in the backward bias condition and gives
rise to thermal rectification. This also explains the parameter dependencies of the rectification ratio
$\eta$ which decreases as temperature, interface coupling and system size increases, and the magnetic field
decreases (for $0 < h <h^*$) since all these factors result in the decrease of spin stiffness at the interface.

\subsection{Thermal rectification and NDTR}
\begin{figure}[htbp]
\begin{center}
\hskip-0.5cm
\includegraphics[width=3.35cm,angle=-90]{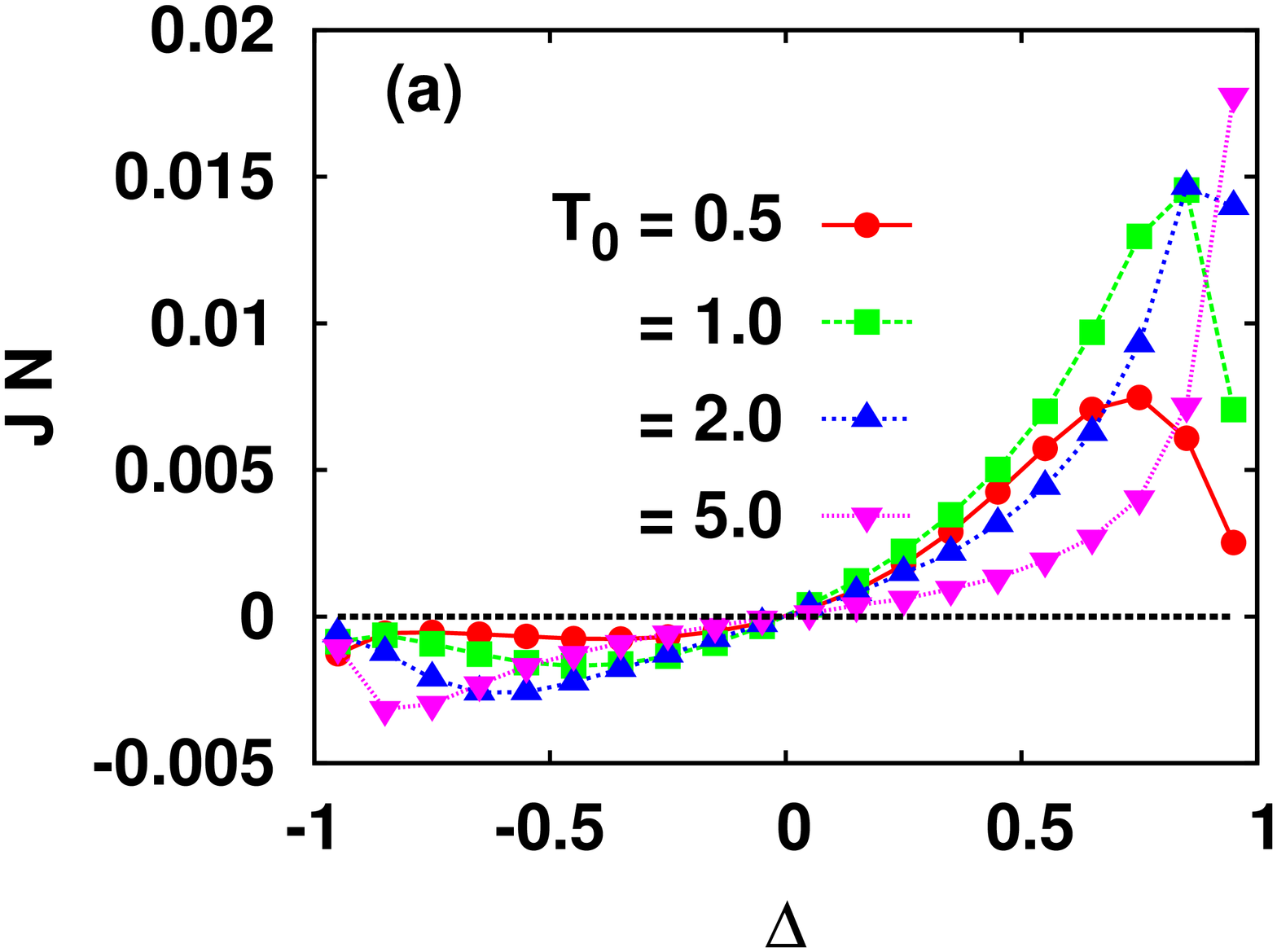}
\hskip-0.5cm
\includegraphics[width=3.35cm,angle=-90]{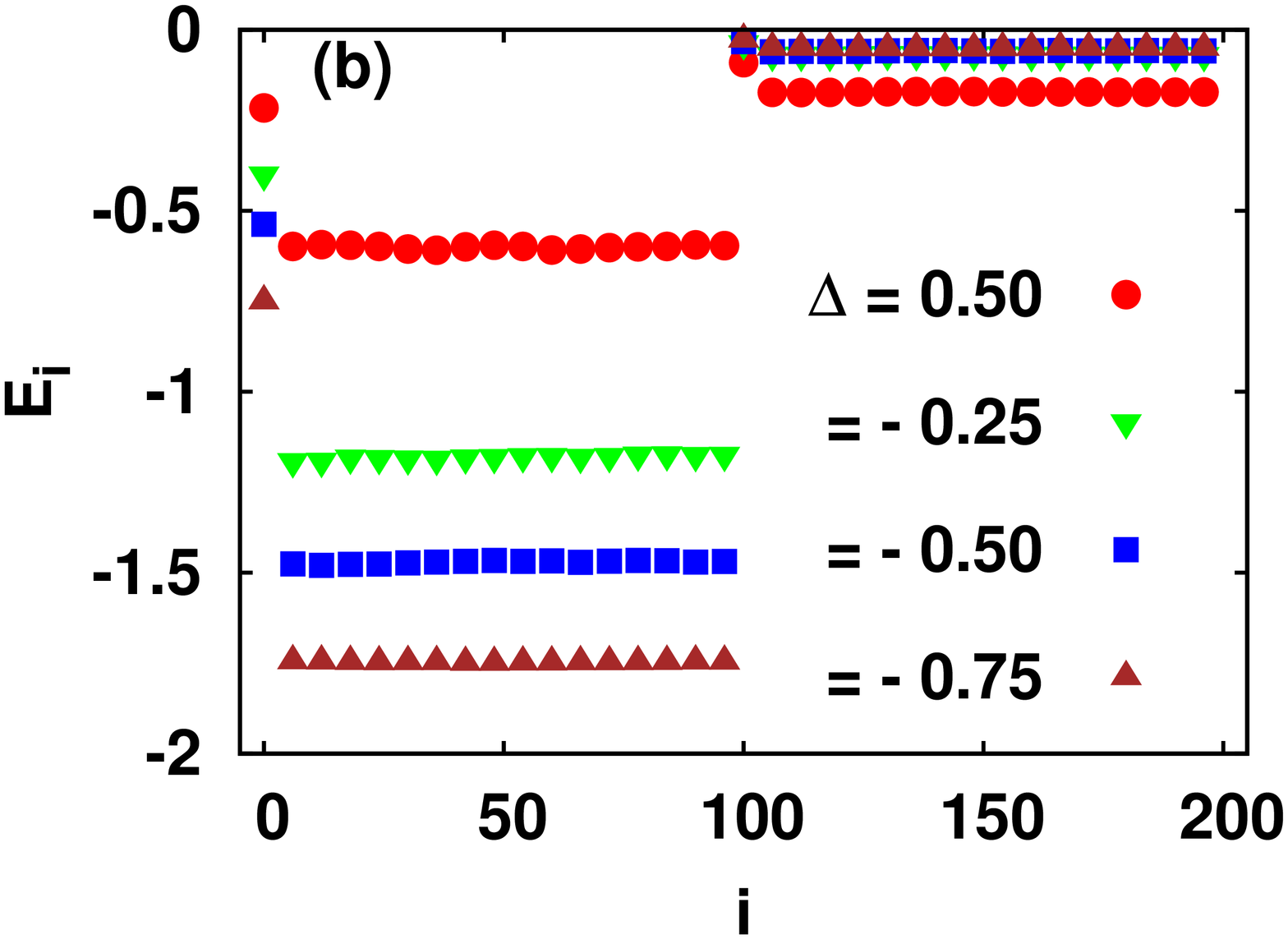} 
\end{center}
\caption{(Color online) (a) Variation of the thermal current $J$ with $\Delta$ for different
average temperatures $T_0 = 0.5, 1.0, 2.0$ and $5.0$. (b) Energy profile of the system for
$\Delta = 0.5$ and $-0.25, -0.5, -0.75$; average temperature $T_0=1.0$. Here $\lambda = 0$ and
all other parameters are the same as in Fig. \ref{fig:Diode}.
}

\label{fig:DiodeZ}
\end{figure}

Next, we set the parameter $\lambda = 0$ and therefore now there is an ``Ising'' type interaction
at the interface of the two segments. We study the thermal current $J$ for different values of the
parameter $\Delta$ keeping all other parameters the same as in the previous case. The result obtained
from simulation is displayed in Fig. \ref{fig:DiodeZ}a. The rectification feature is again seen in this
case - the forward current (for $\Delta > 0$) is appreciably larger in magnitude than the backward
current (for $\Delta < 0$). The energy profiles for different values of $\Delta$ is shown in
Fig. \ref{fig:DiodeZ}b. The rectification effect here can be again explained as in the previous
case.

In addition, careful observation reveals that for certain regions of the $J \sim \Delta$ curves in
Fig. \ref{fig:DiodeZ}a, the current actually decreases as $|\Delta|$ is increased. Thus unlike the
case of $\lambda = 1$, where the magnitude of the current $J$ appears to be a strictly non-decreasing
function of the parameter $\Delta$, for $\lambda = 0$ it is a non-monotonic function. This feature
can be seen in certain ranges for all the four $J \sim \Delta$ curves shown in Fig. \ref{fig:DiodeZ}a.
This is known as the \textit{negative differential thermal resistance} (NDTR). In the following, we
discuss in detail the NDTR feature seen here and try to understand how different factors influence
the emergence of the NDTR region.

%%%%%%%%%%%%%%%%%%%%%%%%%%%% Parameter Dependence of NDTR %%%%%%%%%%%%%%%%%%%%%%%%%%%%%%%%%%%%%%%%%%%%

We focus on the $\Delta < 0$ region of the $J \sim \Delta$ curve and first study the parameter
dependencies of the NDTR feature. Since it is difficult to compare NDTR regimes for different
parameters and system sizes directly (because the current has different typical values), we
define a quantity $\gamma \equiv (J_{m+1} - J_m)/J_m$ which is essentially the change in the
thermal current scaled by the typical value of the current, for two consecutive discrete values
of $\Delta$ belonging to $\{\Delta_m\}$ where $1 \le m \le M$, M being the total number of $\Delta$
values for which the current has been numerically computed. Note that if $|\Delta_{m+1}| > |\Delta_{m}|$
then $\gamma$ is positive for positive differential thermal resistance (PDTR) and negative in the
NDTR regime. Thus $\gamma$ indicates the onset and also the width of the NDTR regime.

\begin{figure}[htbp]
\begin{center}
\hskip-0.5cm
\includegraphics[width=3.2cm,angle=-90]{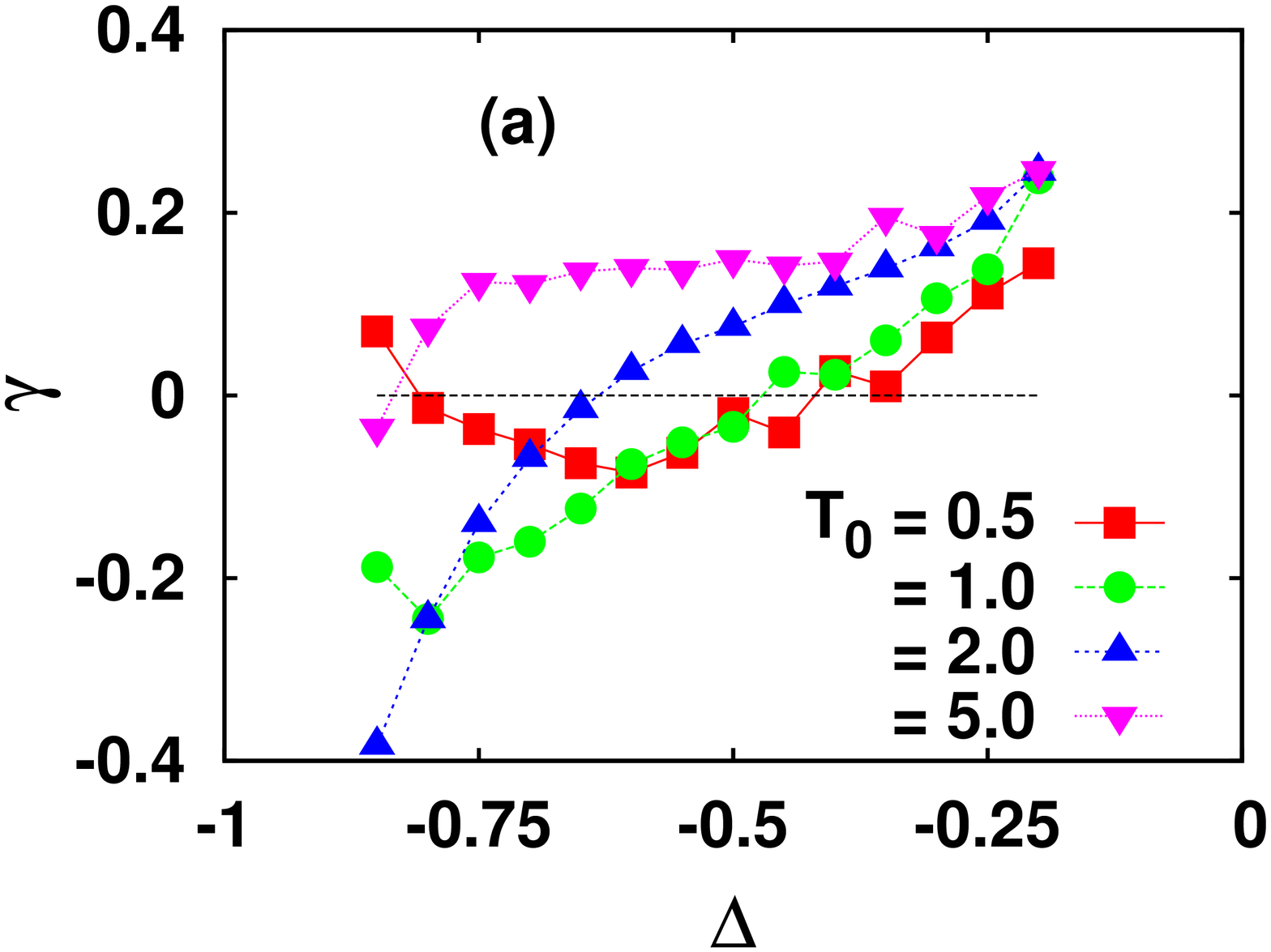}\hskip-0.35cm
\includegraphics[width=3.2cm,angle=-90]{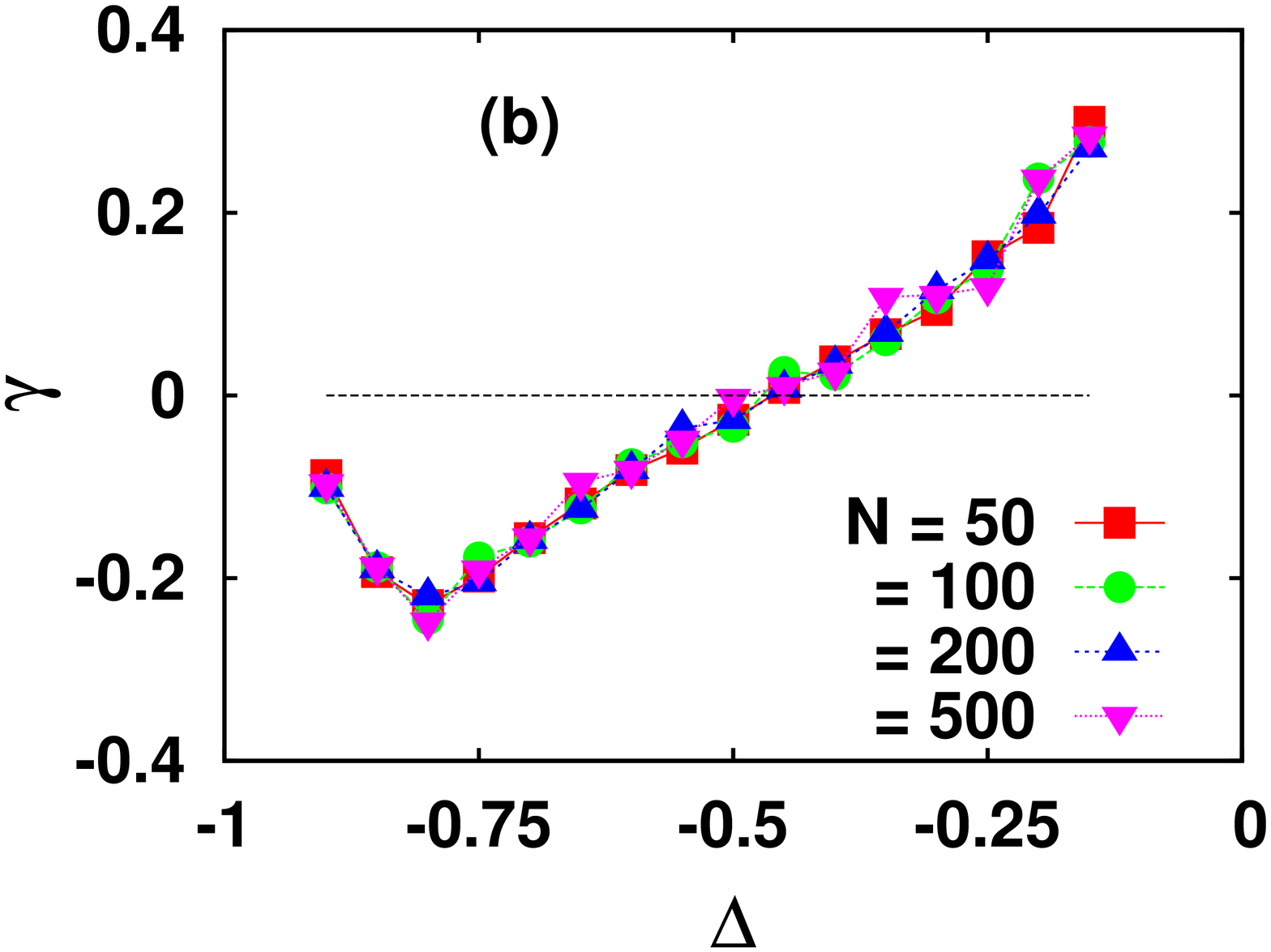}\\
\hskip-0.5cm
\includegraphics[width=3.2cm,angle=-90]{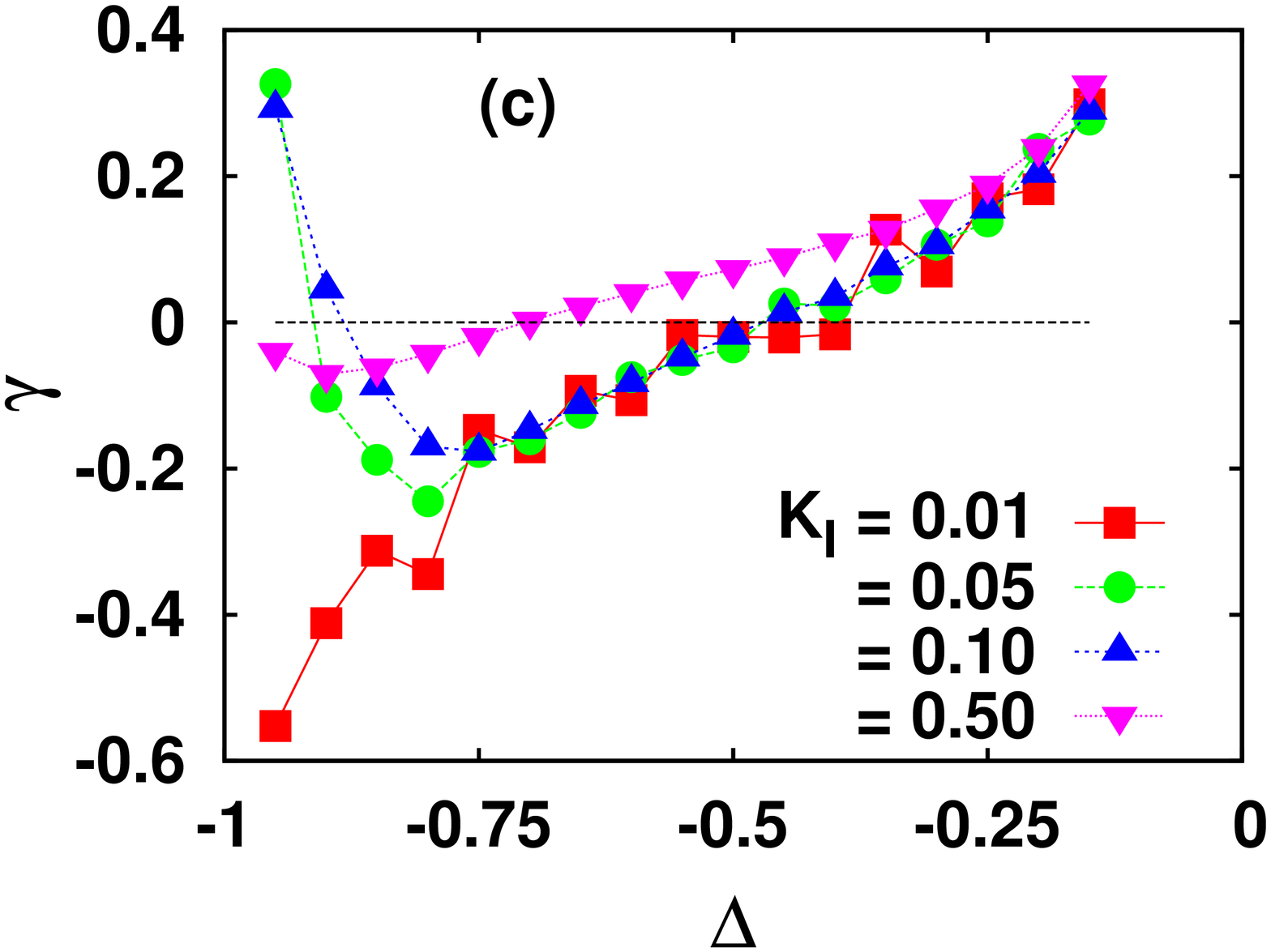}\hskip-0.35cm
\includegraphics[width=3.2cm,angle=-90]{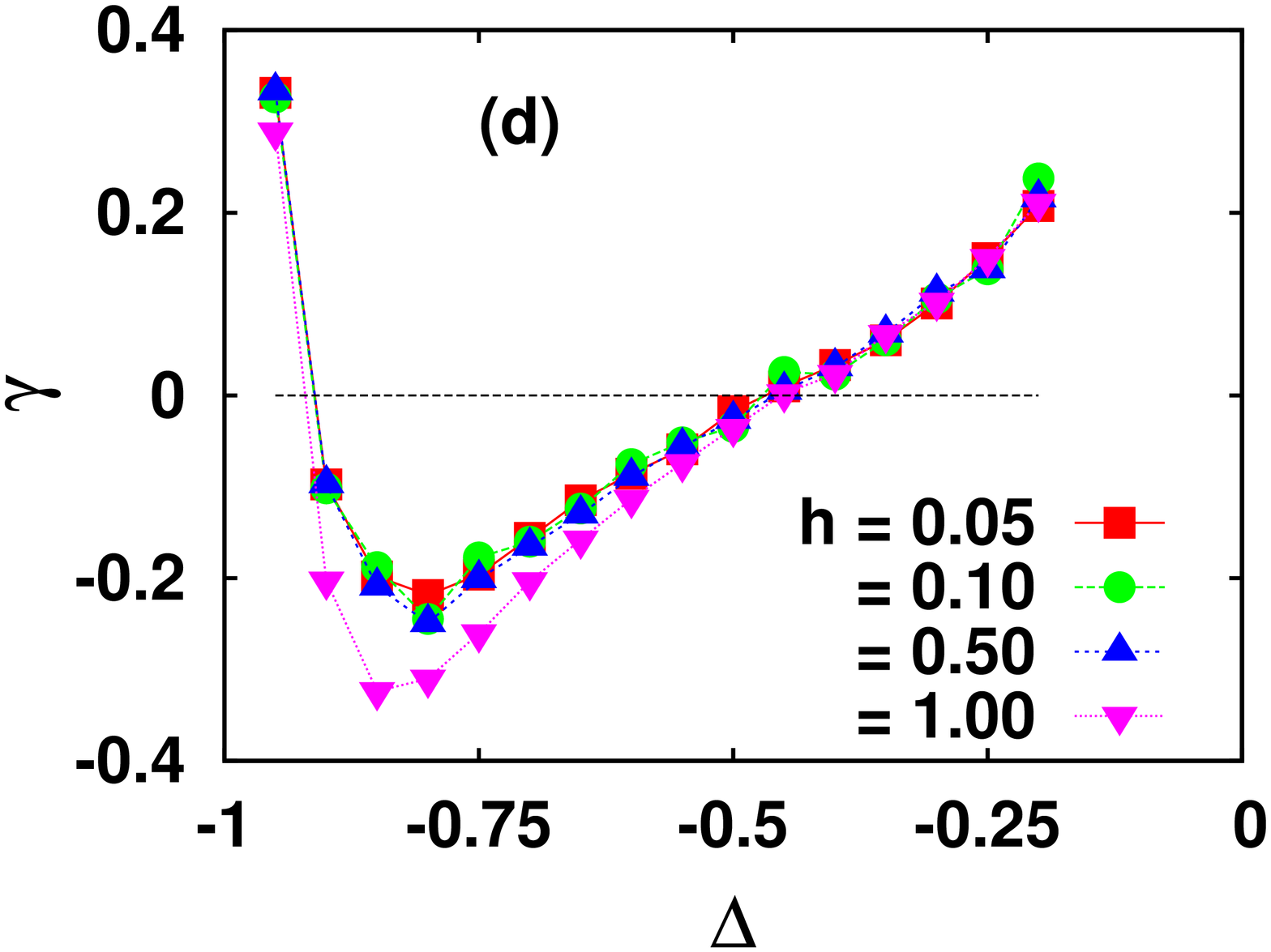}
\end{center}
\caption{(Color online) 
Variation of $\gamma$ with $\Delta$ for different 
(a) average temperature $T_0$ (b) segment size $N$ (c) interface coupling $K_I$ and
(d) magnetic field strength $h$. The values of the parameters are set as $K = 0.5$,
$T_0 = 1.0$, $K_I = 0.05$, $h = 0.1$,  and $N = 100$, and only one of the parameters
is varied in each of the above plots.
}
\label{fig:DataNDTR}
\end{figure}

In Fig. \ref{fig:DataNDTR} we show the quantity $\gamma$ with $\Delta$ for different
parameters $T_0$, $h$, $K_I$ and size $N$. We find that the NDTR regime sets in for
smaller $\Delta$ values at lower average temperature $T_0$ and the NDTR region
vanishes as $T_0$ is increased. It is also found that NDTR is more pronounced
for smaller values of the interface coupling $K_I$. With system size the NDTR regime
remains unaffected which is strikingly different from the result obtained for generic
nonlinear models where the NDTR regime vanishes for larger sizes \cite{origin}. Thus
for a wide range of segment sizes (which differ by an order of magnitude, see Fig.
\ref{fig:DataNDTR}b) the onset of NDTR occurs at the same value of $\Delta$ and there
is no noticeable difference in the onset or the width of the NDTR region. This size
independence of the NDTR here is due to the fact that for the given choice of
parameters the system is very close to the ballistic regime where the thermal current
is independent of the system size. The onset of the NDTR phase for different values of
the magnetic field occurs at the same value of $\Delta$; for larger fields the magnitude
of resistance seems to be slightly larger.

%%%%%%%%%%%%%%%%%%%%%%%%%%%%%%%%%%%%%%%%%%%%%%%%%%%%%%%%%%%%%%%%%%%%%%%%%%%%%%%%%%%%%%%%%%%%%%

Many factors have been considered to be relevant for the occurrence of the NDTR regime and these
have been analyzed in generic phononic systems recently. In phononic systems, the emergence
of the NDTR region has been explained in terms of the mismatch of the phonon bands of the two
particles at the interface \cite{transistor,gate}.
This however has its problems, since it was seen that the band overlap does not depend on the
system size whereas NDTR seems to disappear for larges system sizes \cite{FK-FK}.
In another work ballistic transport has been thought to be responsible for NDTR and it was
shown that an NDTR to PDTR crossover can take place if there is an accompanying ballistic
to diffusive crossover \cite{ballistic_yes}. However a subsequent work showed that NDTR can
emerge even in absence of such crossover of transport processes \cite{ballistic_NO}
%.
Another work suggested that the increasing interface resistance competes with the increasing
temperature gradient and this gives rise to NDTR \cite{origin}. The occurrence of the NDTR was
thought to be the consequence of the nonlinear response of the lattice which causes the interface
resistance to behave nonlinearly for larger thermal gradients. The current increases regularly as
the gradient increases but the interface resistance also increases with the gradient. When the
decrease in the current due to the interface resistance dominates the increases in current due
to the imposed temperature gradient NDTR emerges in such two segment nonlinear systems \cite{origin}.
From Fig. \ref{fig:DiodeZ}b we find that the energy jump for the two dissimilar lattice increases
with the bias. The interface resistance is given as $R_I = \Delta E/J$, where $\Delta E$ is the
energy jump at the interface and $J$ is the current through the system.  
However we do not have any means to independently compute both the interface resistance $R_I$ as well
as the current $J$ and thus it is not clear how $R_I$ varies as the bias $\Delta$ is increased.
In Ref. \cite{transition}, the existence of a critical value for system size $N$ and link interaction
constant $K_I$ was considered above which there is no NDTR and a corresponding phase diagram was
constructed in the $N \sim K_I$ space. Also it was observed that the interaction at the interface is a
crucial factor for NDTR - adding nonlinearity to the link interaction results in the disappearance of
NDTR \cite{transition}. In the following, we seek to resolve and reconcile these issues using our spin
model, and investigate the microscopic mechanism that gives rise to NDTR.

\begin{figure}[htbp]
\begin{center}
\includegraphics[width=4.5cm,angle=-90]{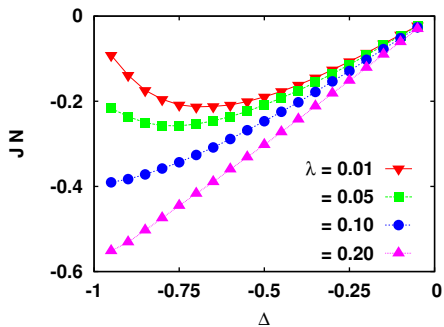}
\end{center}
\caption{(Color online) The variation of thermal current with $\Delta$ in a spatially symmetric
lattice for different values of $\lambda$. The parameters chosen are $K = 1.0$, $K_I = 1.0$,
$h = 1.0$, $T_0 = 1.0$ and segment size $N = 100$. The NDTR regime disappears as $\lambda$ is
increased.
}
\label{fig:lambda}
\end{figure}

To have a better insight let us first understand the difference between the two cases i.e. $\lambda = 0$
and $1$; the former clearly shows NDTR (Fig. \ref{fig:DiodeZ}a) whereas the latter (Fig. \ref{fig:Diode}a)
does not, for the same choice of system parameters. Consider only two spins $\vec S_i$ and
$\vec S_j$ ($j = i \pm 1$) which interact via Eq. \ref{H_I} with $\lambda = 0$.
The equations of motion for the two spins are 
\begin{eqnarray}
\dot S_{i,j}^{x} &=& K_I S_{i,j}^{y} S_{j,i}^z   \nonumber\\
\dot S_{i,j}^{y} &=& - K_I S_{i,j}^{x} S_{j,i}^z \nonumber\\
\dot S_{i,j}^{z} &=& 0.
\label{l=0}
\end{eqnarray}
Thus the equation of motion for both the spins is linear for $\lambda = 0$ whereas it is nonlinear for 
$\lambda = 1$. We have simulated our system for intermediate values of $\lambda$ in the range $0 <
\lambda < 1$ and the results are presented in Fig. \ref{fig:lambda}. As expected we find that for
smaller $\lambda$ values, the NDTR regime appears in the $J \sim \Delta$ curve whereas for larger
values it vanishes, keeping all other parameters the same. Therefore, it is relatively easy to
observe NDTR if the link interaction is linear whereas adding nonlinearity results in the
disappearance of NDTR.
Also it is clear from Eq. (\ref{l=0}) that for $\lambda = 0$ the interface spin stiffness in comparatively
larger (since $\dot S_{i,j}^z = 0$) and restricts energy flow across the interface; this restriction, we
suspect, is responsible for the emergence of NDTR for $\lambda = 0$ and not for $\lambda = 1$, with all system
parameters kept the same. Note that in Fig. \ref{fig:lambda} we have chosen a spatially symmetric lattice
and yet exhibits NDTR which clearly shows that asymmetry is not an essential criterion; NDTR can emerge in
perfectly symmetric systems. Also, since the system is symmetric, it is straightforward to construct the
$J \sim \Delta$ curves shown in Fig. \ref{fig:lambda} for $0 \le \Delta \le 1$.

\begin{figure}[htbp]
\begin{center}
\hskip-0.25cm
\includegraphics[width=3.25cm,angle=-90]{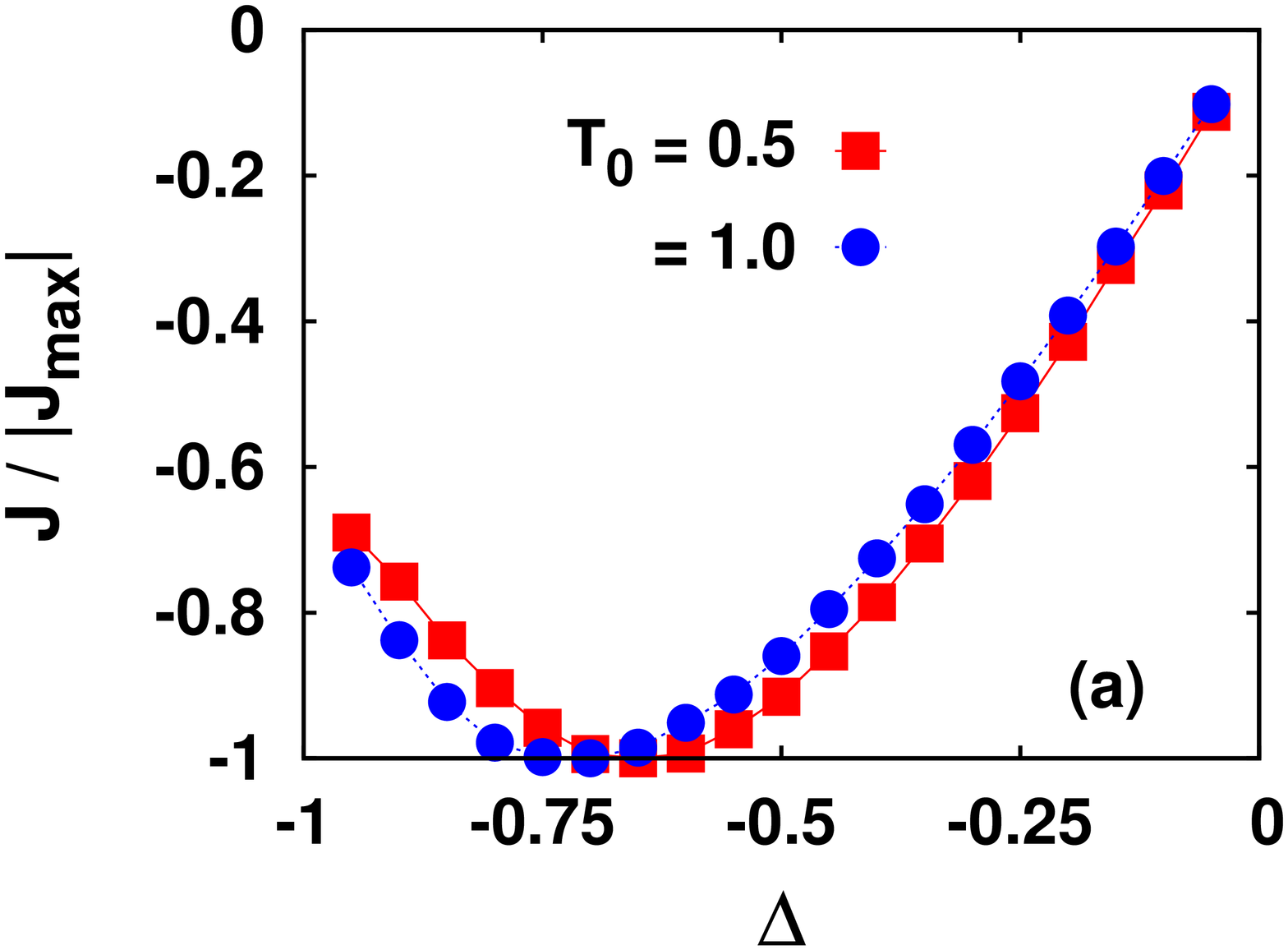}
\hskip-0.35cm
\includegraphics[width=3.25cm,angle=-90]{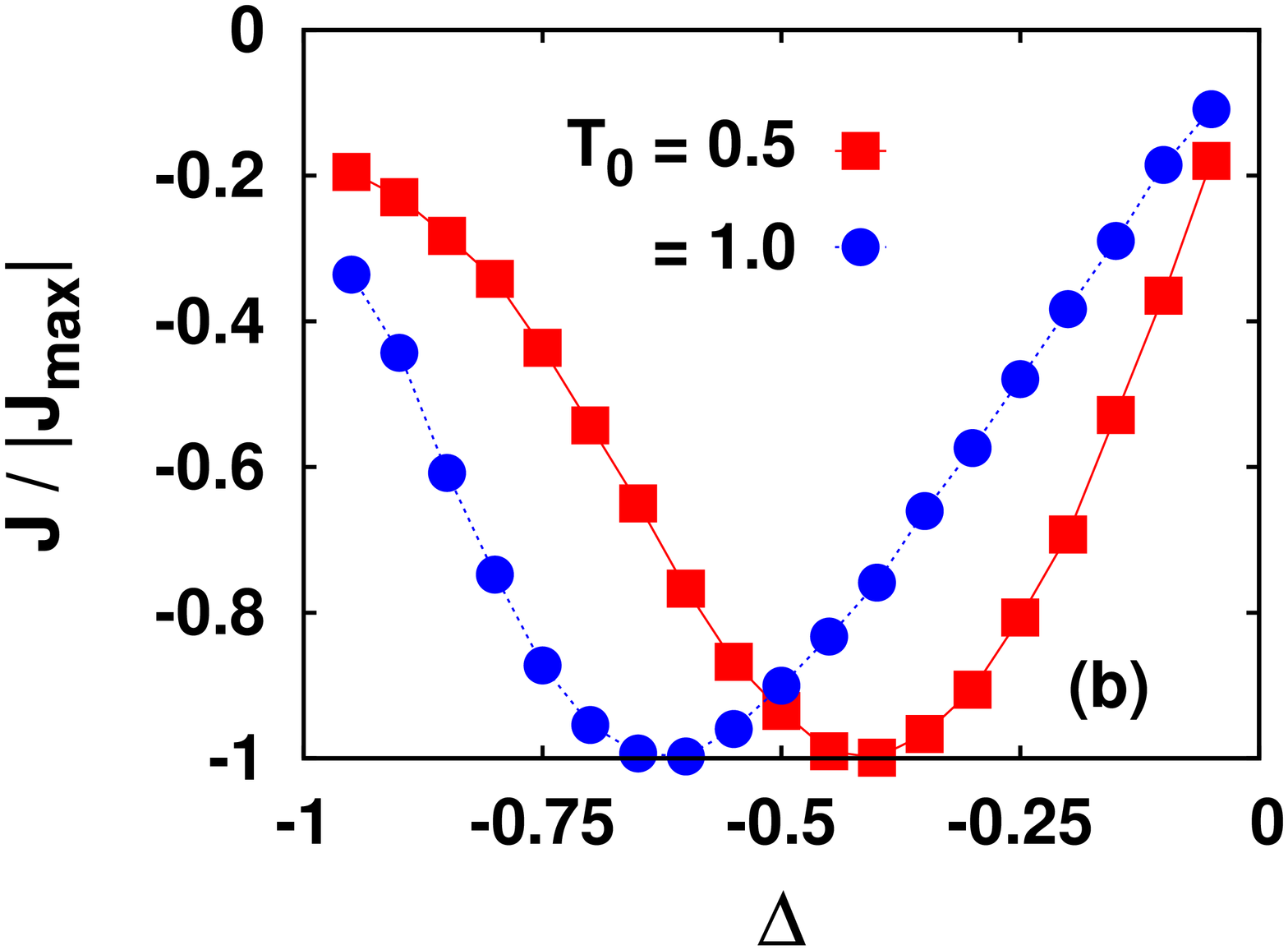}
\end{center}
\caption{
(Color online) The variation of thermal current $J$ with $\Delta$ for  
(a) magnetic field $h = 20.0$ for $\vec S^L_N$ and $\vec S^R_1$ (for all other spins $h = 1.0$)
and other parameters  $K = 1.0$, $K_I = 1.0$ and (b) $K_I = 0.01$ with $K = 1.0$, and $h = 1.0$
for all spins. Here $\lambda = 1$ and the current $J$ has been scaled by the maximum (absolute)
value of the current $|J_{max}|$ for better comparison. Note that the system is homogeneous
(identical spin-spin interaction for all spins in the system) in (a) and segments are spatially
symmetric in (b). Here the segment size $N = 100$.
}
\label{fig:H20K01}
\end{figure}

We speculate that any factor which impedes the energy flow through the link should give rise
to NDTR. If this is true then even for $\lambda = 1$ we should get NDTR by judiciously tuning
system parameters and restricting the current across the link.
Since the average current through the $i$-th bond (between $\vec S_i$ and $\vec S_{i+1}$) 
interacting via $\mathcal{H}_I$ with $\lambda = 1$ (in absence of magnetic fields) can be
expressed as $ J_i = \langle K_I (\vec S_i \times \vec S_{i+1}) \cdot  \vec S_{i+2} \rangle$
\cite{our}, the current $J_i$ can be restricted in either of the two ways - (a) by applying
high local magnetic fields on spins on either side of the bond (this will make $\vec S_i$
almost parallel to $\vec S_{i+1}$ and hence $|\vec S_i \times \vec S_{i+1}| \approx 0$) and
(b) by decreasing interaction strength $K_I$ between the two spins $\vec S_i$ and $\vec S_{i+1}$.
(Recall that the interface resistance is given as $R_I = \Delta E/J$; thus a smaller $J$ across
the interface implies larger resistance.)
These are applied to the interface spins, $\vec S_N^L$ and $\vec S_1^R$, and the results are
displayed in Fig. \ref{fig:H20K01}a and b respectively.
We find that in both cases there  is a clear emergence of the NDTR region. Note that the parameters
used in Fig. \ref{fig:H20K01}a (other than $\lambda$) are the same as that of Fig. \ref{fig:lambda}
except for a stronger magnetic field for spins $\vec S^L_N$ and $\vec S^R_1$. Likewise for
Fig. \ref{fig:H20K01}b, all parameters remain the same as in Fig. \ref{fig:lambda} except
for a lower value of the link interaction strength $K_I$.
Thus in conformity with our proposition, we find that NDTR arises when the energy flow across the link
is suitably restricted even if link interaction is highly nonlinear. This mechanism also consistently
explains all the parameter dependencies for NDTR. For example in large systems (or equivalently at high
temperatures), the transport process approach the diffusive regime and the stiffness of the spins decrease
and this eases energy flow across the link which results in the disappearance of NDTR. However, in our
system we can make NDTR insensitive to system size (see Fig. \ref{fig:DataNDTR}b) by properly tuning other
parameters, such as the local magnetic fields, so that the large spin stiffness at the interface is maintained. 
Thus this simple physical mechanism satisfactorily explains the puzzling issues concerning NDTR.
We find that our results are consistent with a recent analytical work \cite{zia} that also presents
a similar idea using a simple particle hop model where such negative responses can emerge naturally
due to obstruction, a feature typical of nonequilibrium steady states in driven systems.

\section{Conclusion}
\label{summary}
To summarize, we have performed extensive numerical study of a two segment thermally
driven classical Heisenberg spin chain in presence of external magnetic field. The spin
couplings and the applied magnetic fields can in general take different values in the left
and the right segment. The composite system is thermally driven by attaching heat reservoirs
at the two ends. By properly tuning system parameters one can achieve thermal rectification
in this system similar to other nonlinear asymmetric systems. Thus the system behaves as a
good conductor of thermal energy along one direction but restricts the flow of energy in the
opposite direction. The rectification efficiency is found to be controlled by nonlinearity
and spatial asymmetry of the lattice.

The rectification of thermal current in such nonlinear asymmetric two segment lattices
can be physically interpreted in terms of the interface resistance which is found to be
larger in one direction as compared to the other. The rectification efficiency drops as
the system size or the average temperature is increased since a crossover occurs from
ballistic to diffusive transport of energy. The decrease in efficiency $\eta$ is smaller
when the system is closer to the ballistic regime and this can be controlled by tuning the
external magnetic field accordingly. Thus deep inside the ballistic regime the efficiency
is practically independent of the system size or the temperature. At the level of individual
spins, we show that the ease of rotation of the interface spins controls the rectification
seen in this system.

Besides rectification, negative differential thermal resistance can also emerge in this
system where the thermal current decreases as the imposed gradient is increased. This
feature emerges in homogeneous, symmetric and asymmetric systems. The underlying mechanism
for the appearance of NDTR is the restriction of the current through the link connecting the
two segments. This can be tuned by suitably controlling the link interaction $\mathcal{H}_I$,
or other local parameters such as $h$, $K_I$. 
Thus for the emergence of NDTR the crucial features seem to be nonlinearity of the system
under investigation and a suitable mechanism to impede the flow of thermal energy in the
bulk of the system. This physical mechanism is not restricted to the spin system studied
here and should also extend to generic nonlinear systems.

\begin{figure}[htbp]
\begin{center}
\includegraphics[width=4.5cm,angle=-90]{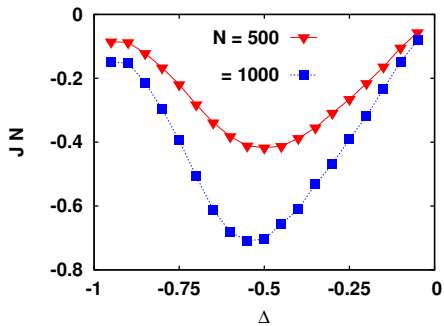}
\end{center}
\caption{(Color online) The variation of thermal current $J$ with $\Delta$ in a homogeneous
lattice with $\lambda = 1$ for segment sizes $N = 500$ and $1000$. The magnetic field $h$ for
the two interface spins $\vec S^L_N$ and $\vec S^R_1$ is $h = - 20.0$ (along the negative $\hat z$
direction) and for all other spins $h = 0.0$; other parameters are $K = 1.0$, $K_I = 1.0$,
$T_0 = 0.5$.
}
\label{fig:LargeNDTR}
\end{figure}

We end with a few brief comments on the experimental realization of NDTR. In some of the previous
studies \cite{ballistic_yes,transition} it was suggested that NDTR will be difficult to implement
in real systems since it is extremely sensitive to interface parameters, temperature and system size.
However in view of the discussion presented here we believe that this should not be, in
principle, very difficult to fabricate. Transport studies in spin systems are of active experimental
interest in recent times \cite{magnon_review}. The classical Heisenberg model is realizable in practice
and chemical compounds which can be mimic classical Heisenberg interactions are known for quite some
time now \cite{jongh,windsor}. 
In such a chemical system the only requirement seems to be a relatively high external magnetic field
at two points close to each other in the bulk which will impede energy flow and give rise to NDTR.
We have also verified that this physical mechanism holds also for large system sizes, as shown in
Fig. \ref{fig:LargeNDTR} and is not a finite size effect.
This proposed experimental setup is a homogeneous symmetric system and does not require any high
precision tuning of the interface properties of the material; only the magnitude of the external
magnetic field needs to be properly controlled (see, for example, the parameter values used in
Fig. \ref{fig:LargeNDTR}). 
Hopefully, with the recent advancement of low dimensional experimental techniques these theoretical
predictions will be verified and lead to the fabrication of devices for efficient thermal management.

\ack 
The author would like to thank P. K. Mohanty for 
helpful suggestions and careful reading of the manuscript.

%%%%%%%%%%%%%%%%%%%%%%%%%%%%%%%%%%%%%%%%%%%%%%%%%%%%%%%%%%%%%%%%%%%%%%%%%%%%%%%%%%%%%%%%%%%%%%%%%%%%%%%%%
%%%%%%%%%%%%%%%%%%%%%%%%%%%%%%%%%%%%%%%%%%%%%%%%%%%%%%%%%%%%%%%%%%%%%%%%%%%%%%%%%%%%%%%%%%%%%%%%%%%%%%%%%

\end{document}